\documentclass[12pt,preprint]{aastex} 
  
\shorttitle{NGC~253 H53$\alpha$ and H92$\alpha$ RRLs}  
\shortauthors{Rodriguez-Rico et al.}  
  
\begin{document}  
  
\title{VLA H53$\alpha$ and H92$\alpha$ line observations of the central region of NGC~253}  
  
\author{C. A. Rodr\'{\i}guez-Rico \altaffilmark{1,2}}  
\email{c.rodriguez@astrosmo.unam.mx}  
  
\author{W. M. Goss\altaffilmark{2}} \email{mgoss@nrao.edu}
  
\author{J.-H. Zhao\altaffilmark{3}} \email{jzhao@cfa.harvard.edu}  
  
\author{Y. G\'omez\altaffilmark{1}} \email{y.gomez@astrosmo.unam.mx}  
  
\author{K. R. Anantharamaiah\altaffilmark{4}}  
  
\altaffiltext{1}{Centro de Radioastronom\'{\i}a y Astrof\'{\i}sica ,  
UNAM, Campus Morelia, Apdo. Postal 3-72, Morelia, Michoac\'an 58089,  
M\'exico.}  
  
\altaffiltext{2}{National Radio Astronomy Observatory, Socorro, NM  
87801}  
  
\altaffiltext{3}{Harvard-Smithsonian Center for Astrophysics, 60  
Garden Street, Cambridge, MA 02138}  
  
\altaffiltext{4}{Raman Research Institute, C.V. Raman Avenue,  
Bangalore, 560 080, India. Deceased 2001, October 29}

\begin{abstract}  
We present new Very Large Array (VLA) observations toward  NGC~253 of the recombination   
line H53$\alpha$ (43~GHz) at an angular resolution of $1\rlap.{''}5 \times 1\rlap.{''}0$.  
The free-free emission at 43~GHz is estimated to be $\sim 100$~mJy,   
implying a star formation rate of $\sim 1.3$~M$_{\odot}$~yr$^{-1}$ in the nuclear region of this starburst galaxy.  
A reanalysis is made for previously reported H92$\alpha$ observations   
carried out with angular resolution of $1\rlap.{''}5 \times 1\rlap.{''}0$
and $0\rlap.{''}36 \times 0\rlap.{''}21$.
Based on the line and continuum emission models used for the $1\rlap.{''}5 \times 1\rlap.{''}0$  
angular resolution observations, the RRLs H53$\alpha$   
and H92$\alpha$ are tracers of the high-density ($\sim 10^5$~cm$^{-3}$) and   
low-density ($\sim 10^3$~cm$^{-3}$) thermally ionized gas components in NGC 253, respectively. 
The velocity fields observed in the H53$\alpha$ and H92$\alpha$ lines   
($1\rlap.{''}5 \times 1\rlap.{''}0$) are consistent. 
The velocity gradient in the central $\sim 18$~pc of the NE component, as  
observed in both the H53$\alpha$ and H92$\alpha$ lines, is in the opposite  
direction to the velocity gradient determined from the CO observations.   
The enclosed virial mass, as deduced from the  
H53$\alpha$ velocity gradient over the NE component, is $\sim 5 \times  
10^6$~M$_{\odot}$ in the central $\sim 18$ pc region.   
The H92$\alpha$ line observations at high angular resolution ($0\rlap.{''}36 \times 0\rlap.{''}21$)  
reveal a larger velocity gradient, along a P.A.$\sim -45^{\circ}$ on the   
NE component, of $\sim 110$~km~s$^{-1}$~arcsec$^{-1}$.  
The dynamical mass estimated using the high angular resolution H92$\alpha$  
data ($\sim 7 \times 10^6$~M$_{\odot}$) supports the  
existence of an accreted massive object in the nuclear region of NGC 253.   
\end{abstract}  


\section{INTRODUCTION}  
NGC~253 is one of the nearest ($\sim 2.5$~Mpc) and brightest starburst  
galaxies, catalogued as an SAB(s)c galaxy \citep{DeV76}. This galaxy  
has an inclination of $\sim 79^{\circ}$ with respect to the  
line-of-sight with major axis located at a position angle (P.A.)  
of 51$^{\circ}$, also containing a bar-like feature tilted by $18^{\circ}$ with  
respect to the major axis \citep{Pen81}. Observations of NGC 253 have  
been carried out in the radio \citep{Tu85,Ul97,Mo02,Mo05,Bo05}  
infrared \citep{En98}, optical \citep{Fo00,Ar95} and X-ray  
\citep{We02} wavelengths. Radio observations, which are not affected by dust  
absorption, are an excellent tool to study the structure and kinematics of the nuclear  
region of NGC~253. Observations in the 21-cm line (Boomsma et  
al. 2005) reveal extra-planar motions of HI that occur at a large  
scale of up to 12~kpc.   
High angular resolution radio continuum observations \citep{Ul97} have revealed a number ($>  
60$) of compact sources in the central 300 pc of this galaxy,  
supporting the scenario of a massive star formation episode occurring  
in the center of NGC~253. Nearly half of these compact continuum  
sources are dominated by thermal radio emission from HII regions  
\citep{Tu85,Ant88,Ul97}.
The radio continuum and radio recombination line  
emission, observed at high angular resolution, have  
been modeled using different density components for the ionized gas  
\citep{Mo05}. The emission models suggest the existence of both low  
($\sim 10^3$~cm$^{-3}$) and high-density ($> 10^4$~cm$^{-3}$) ionized  
gas in the central region of NGC 253. On the other hand, the most  
luminous source (5.79-0.39) is unresolved ($< 1$~pc) at 22 GHz suggesting the  
existence of an AGN in the center of this galaxy \citep{Ul97}.  
Observations of broad H$_2$O maser line emission ($\ge  
100$~km~s$^{-1}$) near this radio continuum source has been invoked as  
further evidence of the presence of a massive object in NGC 253  
\citep{Na95}. \citet{Mo02} modeled the VLA continuum and radio  
recombination line (RRL) emission for the nuclear region of NGC~253  
and favor an AGN as the source responsible for the ionization.  
Observations of hard X-ray emission toward the core of NGC 253  
were also interpreted as evidence of AGN activity \citep{We02}.  
  
The bar-like structure was first observed toward NGC 253 in the  
 near-infrared (NIR), covering the inner $150''$ region of the galaxy  
 \citep{Sco85,Fo92}. The existence of the stellar bar is supported by  
 the observed morphology at optical and mid-infrared frequencies  
 \citep{Fo92,Pi92}. A counterpart of the stellar bar in NGC 253 has  
 been found in CO \citep{Ca88}, HCN \citep{PaT95} and CS \citep{Pe96}.  
Observations in the RRL H92$\alpha$ \citep{An96} at an angular resolution of   
$ 1\rlap.{''}8 \times 1\rlap.{''}0$ reveal a velocity field that is discrepant with  
the CO, CS and HCN observations. \citet{An96} proposed that the  
kinematics observed in the H92$\alpha$ could result from a merger  
of two counter-rotating disks.  
The observed H92$\alpha$ and CO line velocity fields were modeled by  
\citet{Das01} using a bar-like potential for NGC 253, which is in   
reasonable agreement with the observed H92$\alpha$ line velocity field.  
However, this kinematical model can  
only reproduce the velocity field of the CO and CS and  
does not agree with the H92$\alpha$ RRL observations.   
Based on the discrepancy of the CO and the ionized gas kinematics,  
\citet{Das01} proposed that the accretion of a compact object   
($\sim 10^6$~M$_{\odot}$) about $10^7$~years ago could account for the velocity field observed in the H92$\alpha$ RRL.  
\citet{Pa04} observed the CO emission at $3''$ angular resolution  
for the inner region and modeled the kinematics of the molecular gas  
 using a bar potential, concluding that motions of the CO  
gas in the central 150~pc are consistent with a bar potential and   
report evidence of the existence of an inner Lindblad resonance (ILR).  
  
Previous interferometric observations of RRLs have been made at low frequencies 
(e.g. $\sim 8.3$~GHz, H92$\alpha$). VLA observations at $1\rlap.{''}5 \times 1\rlap.{''}0$
angular resolution were used by Anantharamaiah \& Goss~(1996) to study the kinematics of NGC 253; 
Mohan et al. (2002, 2005) used the VLA observations at $1\rlap.{''}5 \times 1\rlap.{''}0$ and $0\rlap.{''}3$
angular resolutions to determine the physical properties of the ionized gas in NGC~253.
In this paper we analyse the kinematics of the ionized gas in the nuclear region of NGC 253
using high frequency RRL observations ($\sim 43$~GHz) and the high angular resolution 
observations (at $0\rlap.{''}3$) in the RRL H92$\alpha$.
Also we use the VLA observations in the RRL H53$\alpha$ and  
the 43~GHz radio continuum, along with previously reported H92$\alpha$ line and 8.3~GHz  
radio continuum observations (Mohan et al.~2005), in order to estimate the physical properties of the ionized gas.
This paper is complementary to the results summarized by Mohan et al.~(2005).
Section~2 presents the observations and  
data reduction, while \S~3 presents the results for the H53$\alpha$ and  
H92$\alpha$ RRLs. In subsection~4.1, a model for the emission of the  
RRLs H53$\alpha$ and H92$\alpha$ as well as for the radio continuum at  
43 and 8.3~GHz is presented. Subsection~4.2 analyzes the kinematics for the  
ionized gas in the center of NGC~253 and \S~5 presents the  
conclusions.  
  
\section{VLA Observations.}  
  
\subsection{H53$\alpha$ line.}  
  
The H53$\alpha$ line ($\nu_{rest}=$43309.4~MHz) was observed in  
the CnD configuration of the VLA on 2003, January 18, 19 and 20.   
We used cycles with integration times of 10~min on NGC 253 and 1~min  
on the phase calibrator J0120-270 ($\sim 0.7$~Jy). Four frequency  
windows (LOs) were used to observe the RRL H53$\alpha$, centered at  
42885.1, 42914.9, 42935.1, and 42964.9~GHz. For each frequency window,  
the on-source integration time was $\sim 2$~hrs, using the mode of 15  
spectral channels with a channel separation of 3.125~MHz ($\sim  
22$~km~s$^{-1}$). The data calibration was carried out for each frequency  
window using the continuum channel, consisting of the central 75\%  
of the band. The flux density scales were determined from  
observations of J0137+331 (3C48; 0.54~Jy). The bandpass response of the  
instrument was corrected using observations of J0319+415 (3C84; $\sim  
7.5$~Jy). The parameters of the observations are summarized in Table~1.  
In order to track reliably the phase variations introduced by   
the troposphere, the calibration of the data was performed correcting for   
the phases in a first step and subsequently correcting for both amplitude and phase.   
The line data were further calibrated using the solutions obtained by  
self-calibrating the continuum channel of each frequency window. The  
radio continuum images were obtained by combining the continuum  
channels of each frequency window using the task DBCON from AIPS, and  
the self-calibration method was also applied to this combined data.  
The H53$\alpha$ line cubes and the 43 GHz continuum image were made  
using a natural weighting scheme and then convolved to obtain a  
Gaussian beam of $1\rlap.{''}5 \times 1\rlap.{''}0$  
(P.A.$=0^{\circ}$). The combination of the different frequency  
windows was made following a similar method to that used for the  
H53$\alpha$ line observed toward M82 \citep{RoR04}: (1) the line data  
from each frequency window were regridded in frequency using the GIPSY  
reduction package, (2) before combining the four LOs into a single  
line cube, the continuum emission was subtracted for each frequency  
window using the AIPS task IMLIN with a zero order polynomial fit based on the line free channels, and   
(3) the four line cubes (after subtraction of the continuum) were   
combined into a single line cube.   
The total line bandwidth, after combining all the windows, is about  
150~MHz (1000~km~s$^{-1}$). The line data cube was Hanning-smoothed  
using the task XSMTH in AIPS to reduce the Gibbs effect and the final  
velocity resolution is $\sim 44$~km~s$^{-1}$.  
  
\subsection{H92$\alpha$ line.}  
  
We also present previously reported observations of the RRL H92$\alpha$   
($\nu_{rest}=$8309.4~MHz) at $1\rlap.{''}5 \times 1\rlap.{''}0$, P.A.$=0^{\circ}$ (Anantharamaiah et al.~1996)  
and $0\rlap.{''}36 \times 0\rlap.{''}21$, P.A.$=-3^{\circ}$ (Mohan et al.~2005, 2002 ) angular resolutions toward NGC 253.  
The H92$\alpha$ RRL images at   
$1\rlap.{''}5 \times 1\rlap.{''}0$ angular resolution were produced by combining  
observations carried out in the B (August 31 and Sept 01, 1990),   
C (May 14 and 23, 1988) and D (July 01 and 19, 1988) configurations of the VLA.  
In order to obtain the same HPFW beam as  
the H53$\alpha$ line cube, the H92$\alpha$ line cube was made using these   
'B+C+D' combined data applying a natural weighting scheme.   
Anantharamaiah \& Goss (1996) have already used these 'B+C+D' combined observations   
to analyze the kinematics of the ionized gas in the central $10''$ of NGC 253 with  
an angular resolution of $1\rlap.{''}8 \times 1\rlap.{''}0$, P.A.$=10^{\circ}$.  
The higher angular resolution ($0\rlap.{''}36 \times 0\rlap.{''}21$, P.A.$=-3^{\circ}$) observations   
of the H92$\alpha$ line toward NGC~253 were made with the VLA in the A array   
(July 9 and 12, 1999) and have been previously reported by \citet{Mo02}.  
Recently, \citet{Mo05} used these H92$\alpha$ data along with observations in  
the RRLs H75$\alpha$ and H166$\alpha$ data to model the RRL and the  
radio continuum emission in order to determine the physical parameters  
of the ionized gas. In this paper we use these high angular resolution observations  
to study the kinematics of the ionized gas in the nuclear $5''$ region of NGC~253.  
Because of the different spectral line grid of the H92$\alpha$ high angular resolution observations and   
the H92$\alpha$ 'B+C+D' data, a combined dataset 'A+B+C+D' was not produced.  
  
For the H92$\alpha$ line observations, the phase calibrator   
was J0118-216 and the bandpass calibrator was J2253+161.  
A spectral mode with 31 channels was used.   
The continuum images were obtained by averaging the data in the central  
75\% of the total band.   
The continuum data were processed using  standard calibration   
and self-calibration procedures.  
The calibration and self-calibration used for the continuum data  
were then applied to the line data.  
All the images were made in the AIPS environment.   
The line images were Hanning-smoothed to reduce the Gibbs  
effect and the velocity resolution is 56.4~km~s$^{-1}$.  
Further observational details are summarized by \citet{Mo02} and \citet{Mo05}.  
  
\section{RESULTS}  

\begin{figure}[!ht]  
\plotone{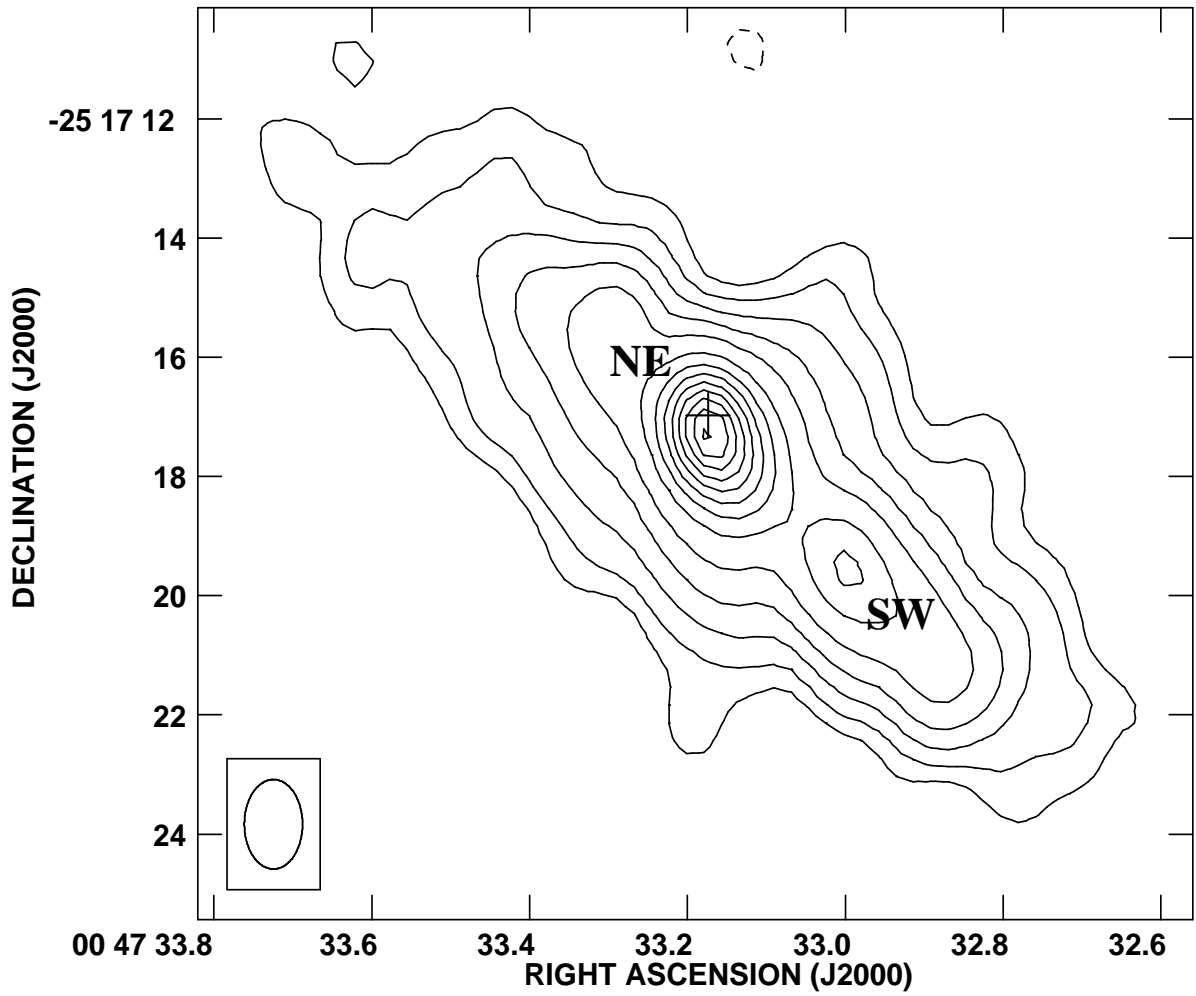}  
\caption{Radio continuum image of NGC~253 at 43~GHz obtained using the VLA. Contour levels  
are drawn at $-3$, 3, 6, 12, 24, 48, 96, 144, 192, 240, 288, 336, 384,  
and 432 times the rms of 0.2~mJy~beam$^{-1}$. The cross shows the  
position of the compact source 5.79-39.0 \citep{Ul97}. The angular  
resolution is $1\rlap.{''}5 \times 1\rlap.{''}0$, P.A.$=0^{\circ}$.}  
\label{f1}
\end{figure}

Figure~\ref{f1} shows the radio continuum emission of NGC~253 at  
43~GHz with an angular resolution of $1\rlap.{''}5 \times 1\rlap.{''}0$, 
P.A.$=0^{\circ}$ ($1'' \simeq 12$~pc).   
The integrated 43 GHz continuum flux density is $360 \pm 20$~mJy, obtained by integrating the flux 
density over the nuclear $30''$ region using the task IRING in AIPS.
The radio continuum image at 43~GHz shows two radio continuum   
components, NE and SW, in addition to extended emission (see Figure~1).   
The continuum peak position of the NE component, $\alpha$(J2000)$=00^{h}47^{m}33\rlap.{^{s}}17 \pm 0\rlap.{^{s}}01$, 
$\delta$(J2000)$=-25^{\circ} 17' 17\rlap.{''}4 \pm 0\rlap.{''}1$, coincides within $0\rlap.{''}2$ with the position of  
the compact source 5.79-39.0 \citep{Ul97}.

\begin{figure}[!ht]  
\plotone{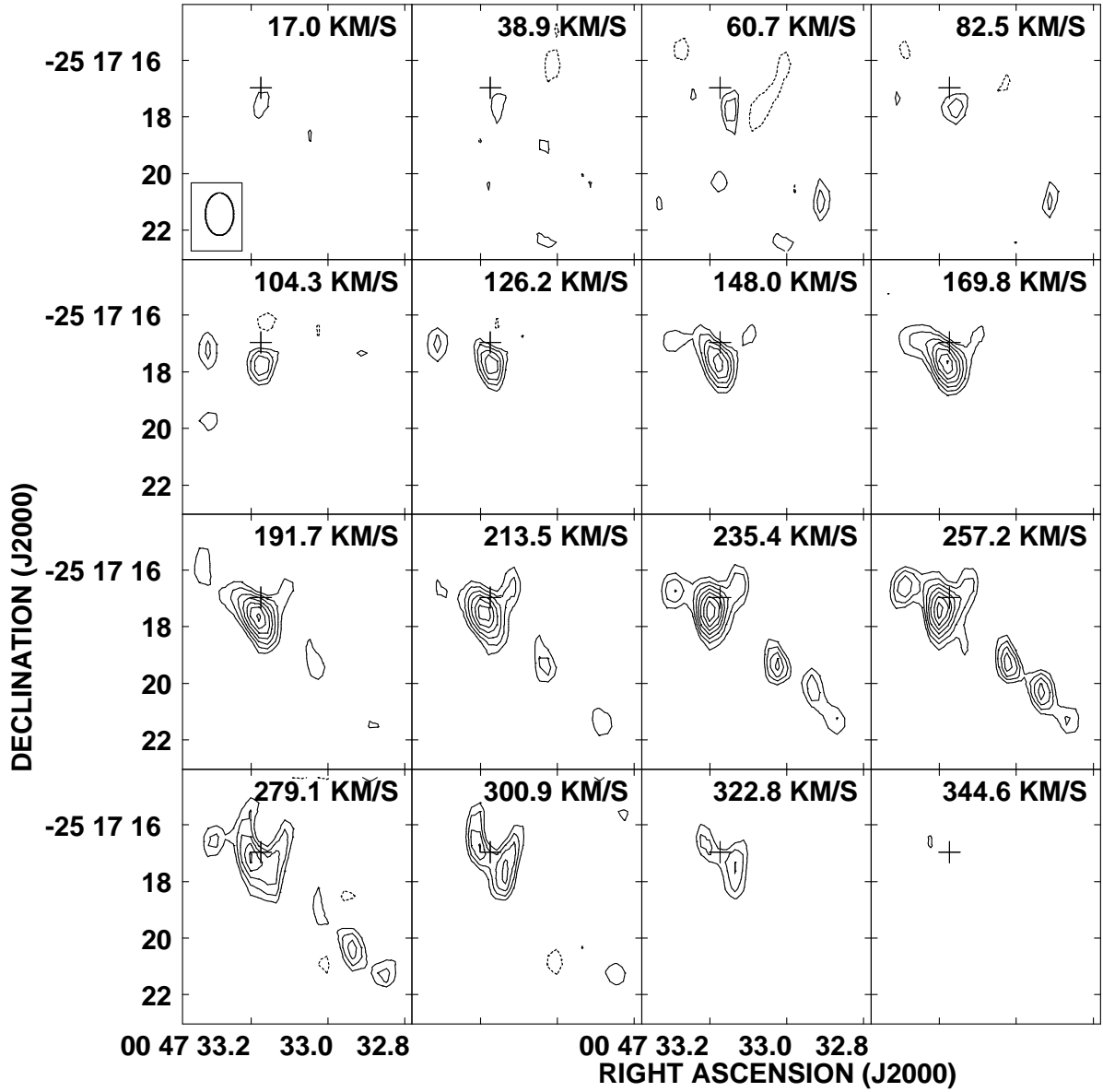}  
\caption{ Channel images of the H53$\alpha$ line emission toward  
NGC~253 obtained using the VLA. Contours are $-3$, 3, 4, 5, 6, 7, 8, 9 and 10 times  
0.65~mJy~beam$^{-1}$, the rms noise. The cross shows the position of  
the compact source 5.79-39.0 \citep{Ul97}. The synthesized beam  
($1\rlap.{''}5 \times 1\rlap.{''}0$ FWHM, P.A.$=0^{\circ}$) is shown in the first panel.  
The central heliocentric velocity is given above for each image.}  
\label{f2}
\end{figure}

\begin{figure}[!ht]
\plotone{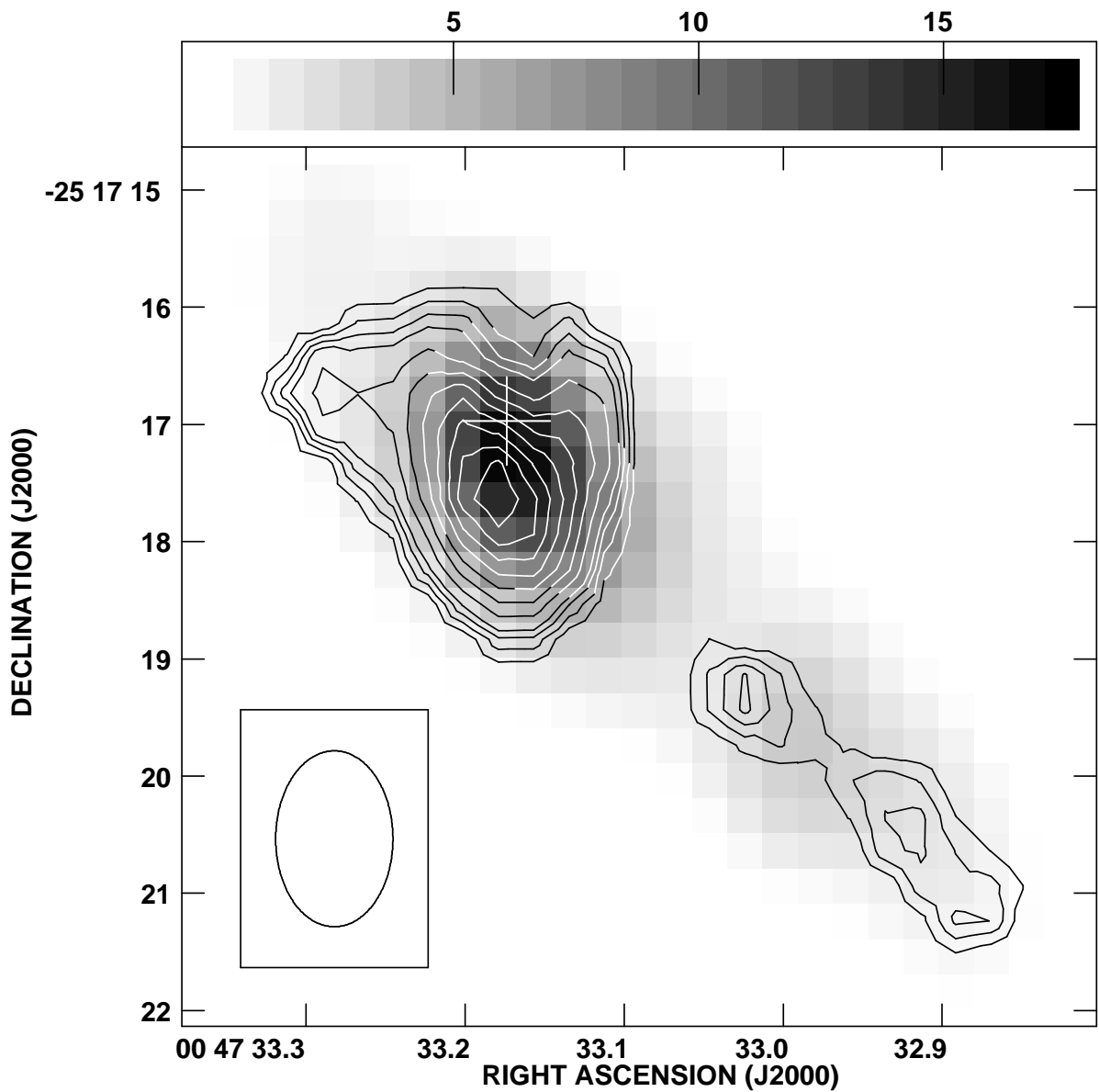}  
\caption{ The velocity integrated (moment 0) of the H53$\alpha$ line  
emission (contours) superposed on the moment 0 of the H92$\alpha$ line  
emission (gray scale) toward NGC 253. Contour levels are drawn at 5, 10, 15, 20, 30, ..., 90\% the peak line emission of   
1.16~Jy~beam$^{-1}$~km~s$^{-1}$. The gray scale (moment 0 of H92$\alpha$) covers the range $0.18-18$~Jy~beam$^{-1}$~km~s$^{-1}$.  
The cross (+) shows the position of the compact source 5.79-39.0  
\citep{Ul97}. The HPFW, for both the H53$\alpha$ and the  
H92$\alpha$ line images is $1\rlap.{''}5 \times 1\rlap.{''}0$, P.A.$=0^{\circ}$.}  
\label{f3}  
\end{figure}

Figure~\ref{f2} shows the H53$\alpha$ velocity-channel images of
NGC~253 with an angular resolution of $1\rlap.{''}5 \times 1\rlap.{''}0$ (P.A.$=0^{\circ}$). 
The H53$\alpha$ line emission is detected toward both  
the NE and SW continuum components above a 3$\sigma$ level ($\sim  
2$~mJy). The ionized gas is observed in the H53$\alpha$ line at  
heliocentric velocities that range from $17$ to $345$~km~s$^{-1}$.  
The velocity-integrated H53$\alpha$ line emission (moment 0) is shown  
in Figure~\ref{f3} superposed on the moment 0 of the H92$\alpha$ line.   
There is good correspondence between the integrated  
line emission of the RRLs H53$\alpha$ and H92$\alpha$.  
In addition, the peak position of the integrated H53$\alpha$ line emission is  
in agreement with the peak position of the 43~GHz radio continuum image.  
In the H53$\alpha$ line images, both the NE and SW components are  
spatially resolved only along the major axis.

\begin{figure}[!ht]  
\plotone{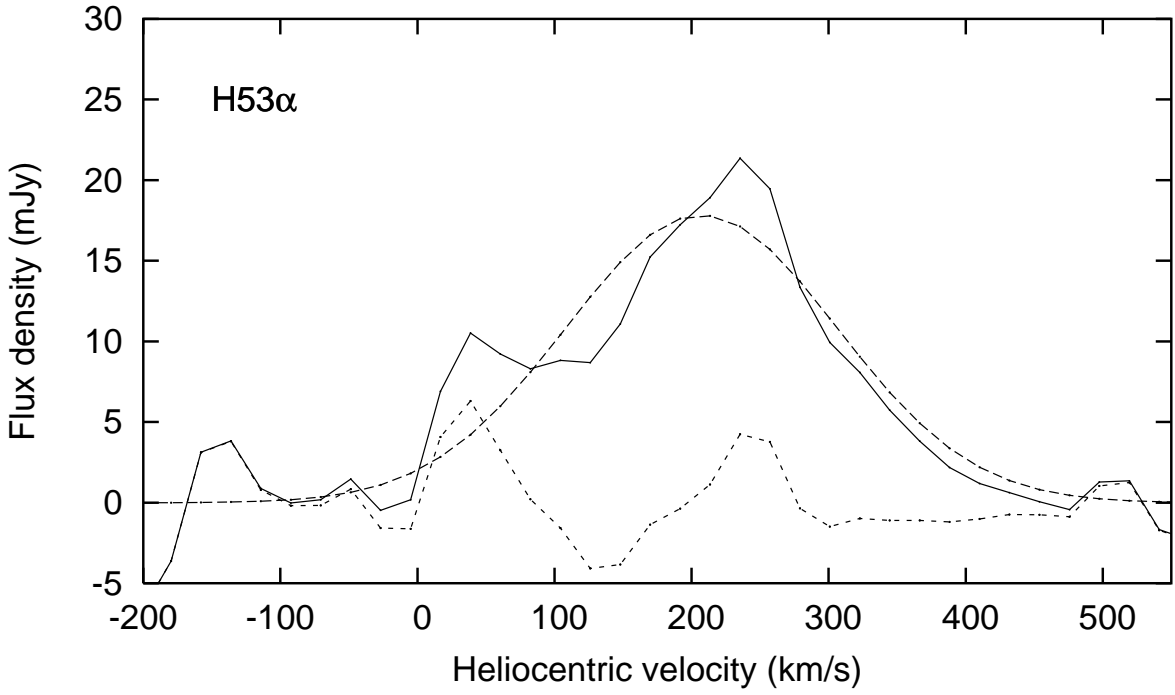}  
\caption{ Total integrated H53$\alpha$ spectrum from NGC 253, obtained by integrating over the central $10''$.   
The thick line shows the data, the dashed line shows the Gaussian fit and the dash-dotted line  
shows the residuals to this fit.}  
\label{f4}  
\end{figure}

Figure~\ref{f4} shows the H53$\alpha$ line spectrum integrated over the central $10''$ region of NGC~253.
By fitting a Gaussian, the estimated central heliocentric velocity is 
$210 \pm 10$~km~s$^{-1}$, the FWHM of  
the line is $230 \pm 20$~km~s$^{-1}$ and the peak line flux density is  
$21 \pm 2$~mJy. The resulting fit is shown in  
Figure~\ref{f4} along with the residuals to the fit. The  
central velocity is in agreement with previous estimates in the optical   
($225 \pm 5$~km~s$^{-1}$; Arnaboldi et al. 1995) and IR   
($230 \pm 10$~km~s$^{-1}$; Prada, Guti\'errez \& McKeith 1998).   
The velocity integrated H53$\alpha$  
line flux density determined from our observations is $0.69 \pm 0.09  
\times 10^{-20}$~W~m$^{-2}$, in agreement with the previous  
measurement of $0.94 \pm 0.38 \times 10^{-20}$~W~m$^{-2}$ derived from  
single-dish observations \citep{Pu97}. A Gaussian function was also  
used to determine the characteristics of the spectra obtained by  
integrating over the NE and SW regions. Table~2 lists the results  
for the total integrated H53$\alpha$ line emission profile, as well as  
for profiles that correspond to the NE and SW  
components. The values listed for the H53$\alpha$ line are peak  
flux density $S_L$, FWHM, the heliocentric velocity V$_{Hel}$ and the velocity integrated  
H53$\alpha$ line emission.  
  
\begin{figure}[!ht]  
\plotone{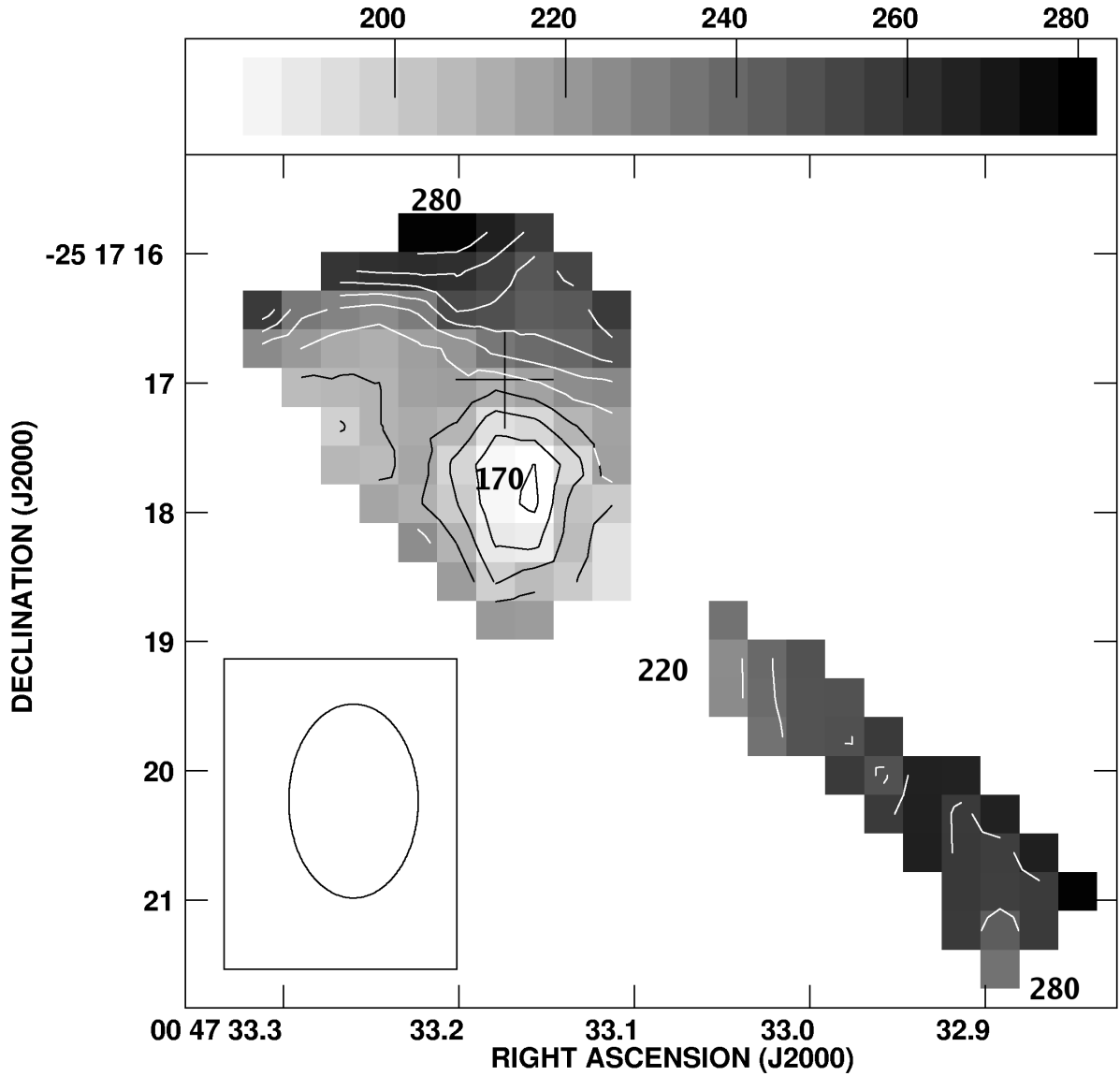}  
\caption{ Velocity field observed in the H53$\alpha$ line from NGC  
 253. Contour levels are the heliocentric velocities at 180, 190, 200, 210, 220, 230, 240, 250, 260, and 270~km~s${^{-1}}$. 
The gray scale ranges from 170 to 280~km~s${^{-1}}$. The cross (+) shows the position of the compact  
 source 5.79-39.0 \citep{Ul97}. The HPFW is  $1\rlap.{''}5 \times 1\rlap.{''}0$, P.A.$=0^{\circ}$.}  
\label{f5}  
\end{figure}

\begin{figure}[!ht]  
\plotone{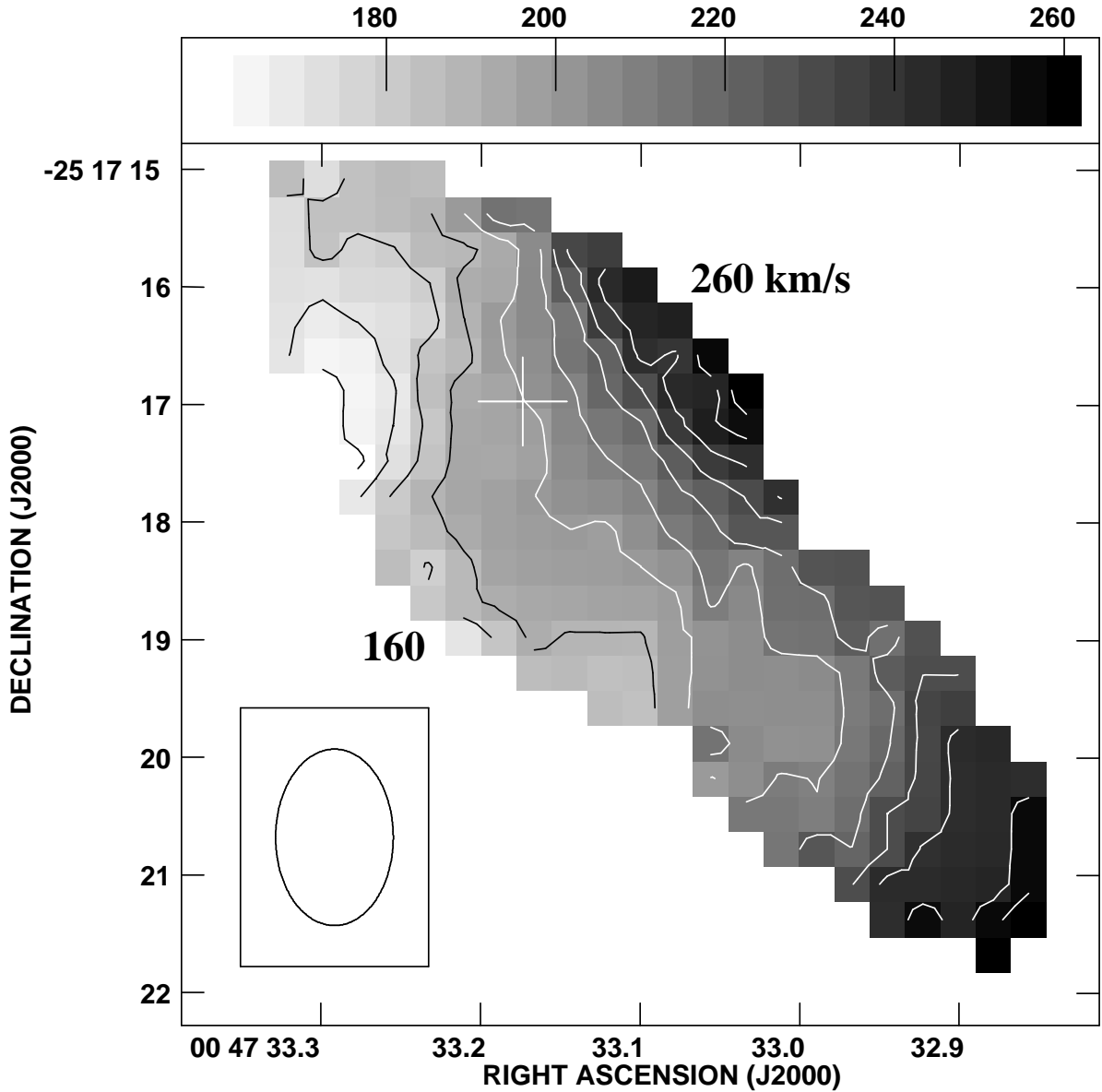}  
\caption{ The H92$\alpha$ velocity field of NGC~253 \citep{An96}, is shown in contours superposed on the gray scale image   
of the same H92$\alpha$ velocity field. Contour levels are drawn for the heliocentric velocities  
of the ionized gas from 160 to 260~km~s$^{-1}$ in steps of 10~km~s$^{-1}$ and the gray scale covers the  
same heliocentric velocity range. The cross (+) shows the position  
of the compact source 5.79-39.0 \citep{Ul97}. The HPFW is  
$1\rlap.{''}5 \times 1\rlap.{''}0$, P.A.$=0^{\circ}$. }  
\label{f6}  
\end{figure}

Figure~\ref{f5} shows the velocity field (moment 1) of the  
H53$\alpha$ line at an angular resolution of $1\rlap.{''}5 \times 1\rlap.{''}0$ (P.A.$=0^{\circ}$).    
Figure~\ref{f6} shows the H92$\alpha$ velocity field made using   
the 'B+C+D' data of \citet{An96} at the same angular resolution.  
The velocity field of the ionized gas as observed in the   
H53$\alpha$ line agrees with observations of the RRL H92$\alpha$   
(see section 4.2 for a detailed comparison).   
In the NE component of NGC 253 the red-shifted gas is observed toward the NW   
and the blue-shifted gas toward the SE.  
In the region located S of the radio continuum peak, there is a blue-shifted component that is more apparent
in the H53$\alpha$ line than in the H92$\alpha$ line. A detailed comparison between the line profiles of the H53$\alpha$ and 
H92$\alpha$ in this region shows that the H53$\alpha$ line is broader than the H92$\alpha$ line by $\sim 50$~km~s$^{-1}$ and 
there is a relative velocity shift between these two RRLs of $\sim 30$~km~s$^{-1}$.
In the elongated SW component, the red-shifted gas is located at the SW  
 and the blue-shifted gas is at the NE.  
The velocity gradient was measured, at $1\rlap.{''}5 \times 1\rlap.{''}0$ angular resolution,  
along the major (P.A.$= 52^{\circ}$) and nearly along the minor (P.A.$= -45^{\circ}$) axis  
for both RRLs the H53$\alpha$ and H92$\alpha$. 
The H53$\alpha$ velocity gradient along the major axis of NGC 253 (measured over the NE component)   
is $12 \pm 3$~km~s$^{-1}$~arcsec$^{-1}$, comparable to the   
corresponding H92$\alpha$ velocity gradient ($\sim 11$~km~s$^{-1}$~arcsec$^{-1}$). 
The velocity gradients measured in the RRLs H53$\alpha$ and H92$\alpha$   
(both at $1\rlap.{''}5 \times 1\rlap.{''}0$) along the P.A.$=-45^{\circ}$
are $42 \pm 8$~km~s$^{-1}$~arcsec$^{-1}$ and $24 \pm 2$~km~s$^{-1}$~arcsec$^{-1}$, respectively.
 
\begin{figure}[!ht]  
\plotone{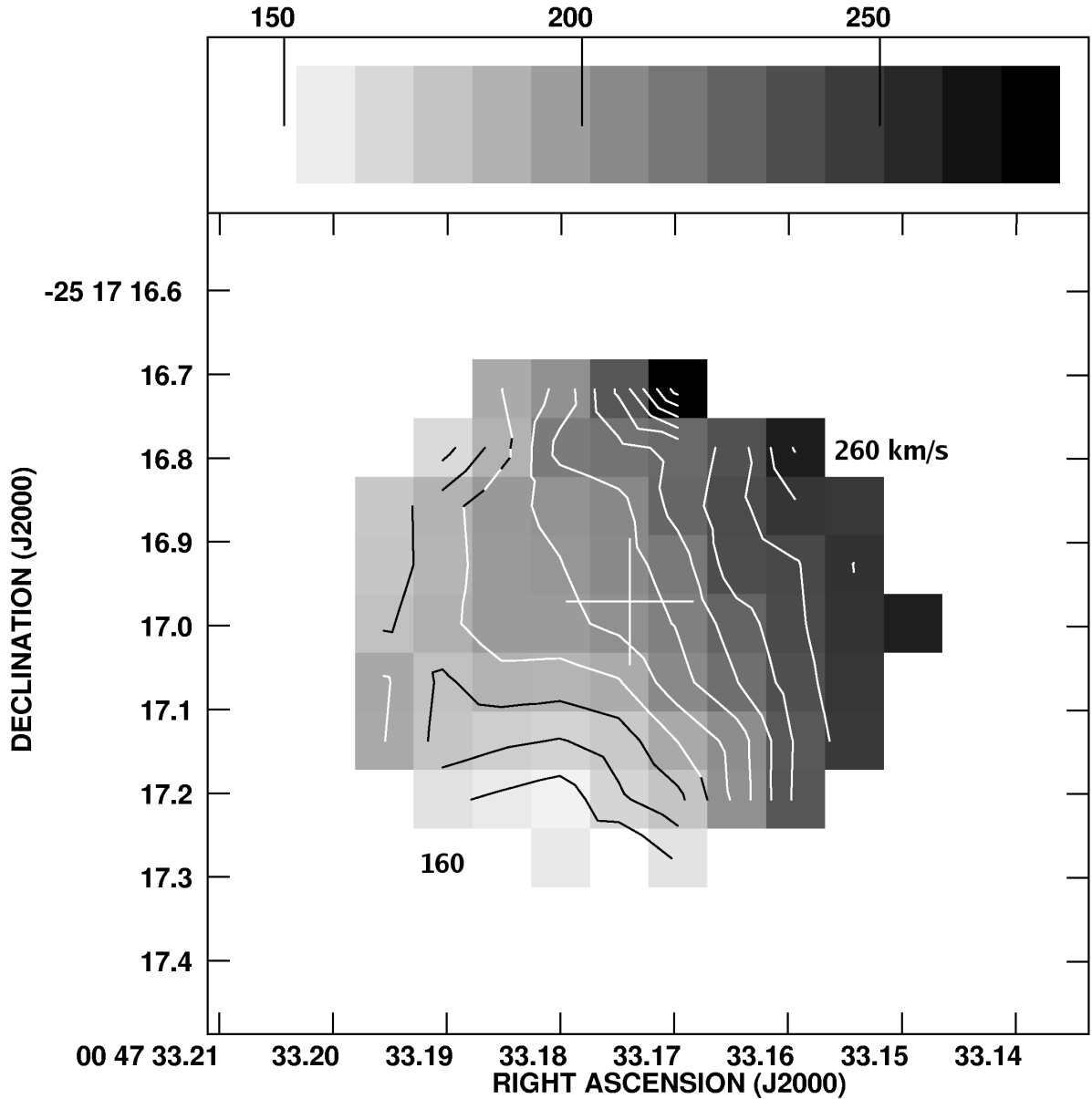}  
\caption{ High angular resolution ($0\rlap.{''}36 \times 0\rlap.{''}21$, P.A.$=-3^{\circ}$)  
heliocentric velocity field (moment 1) image of the H92$\alpha$ line emission  
toward NGC 253 obtained using the VLA A array data. Contour levels are drawn at 160, 170, ..., 260  
km~s$^{-1}$. The cross (+) shows the position of the compact source  
5.79-39.0 \citep{Ul97}. }  
\label{f7}  
\end{figure}

Figure~\ref{f7} shows the H92$\alpha$ velocity field at an angular resolution  
of $0\rlap.{''}36 \times 0\rlap.{''}21$ (P.A.$= -3^{\circ}$). 
At this angular resolution, the H92$\alpha$ line emission is detected only  
toward the NE component with an angular size of $\sim 0\rlap.{''}6$ ($7$~pc).  
Based on these H92$\alpha$ data, a larger velocity gradient of $110 \pm 20$~km~s$^{-1}$~arcsec$^{-1}$   
is measured  along a P.A.$\simeq -45^{\circ}$.  
This velocity gradient is about a factor of four larger than the velocity gradient estimated  
using the lower angular resolution H92$\alpha$ observations ($1\rlap.{''}5 \times 1\rlap.{''}0$).  
The lower velocity gradient measured in the low angular resolution image of H92$\alpha$   
($1\rlap.{''}5 \times 1\rlap.{''}0$) is due to a beam dilution effect.  
By convolving the high angular resolution data of the H92$\alpha$  
with a Gaussian beam of $1\rlap.{''}5 \times 1\rlap.{''}0$ angular resolution,  
the velocity gradient measured along the P.A.$=-45^{\circ}$ is consistent with  
the lower angular resolution H92$\alpha$ data.  
  
\section{DISCUSSION}  
  
\subsection{Models for the radio continuum and recombination line emission}  
  
\begin{figure}[!ht]  
\plotone{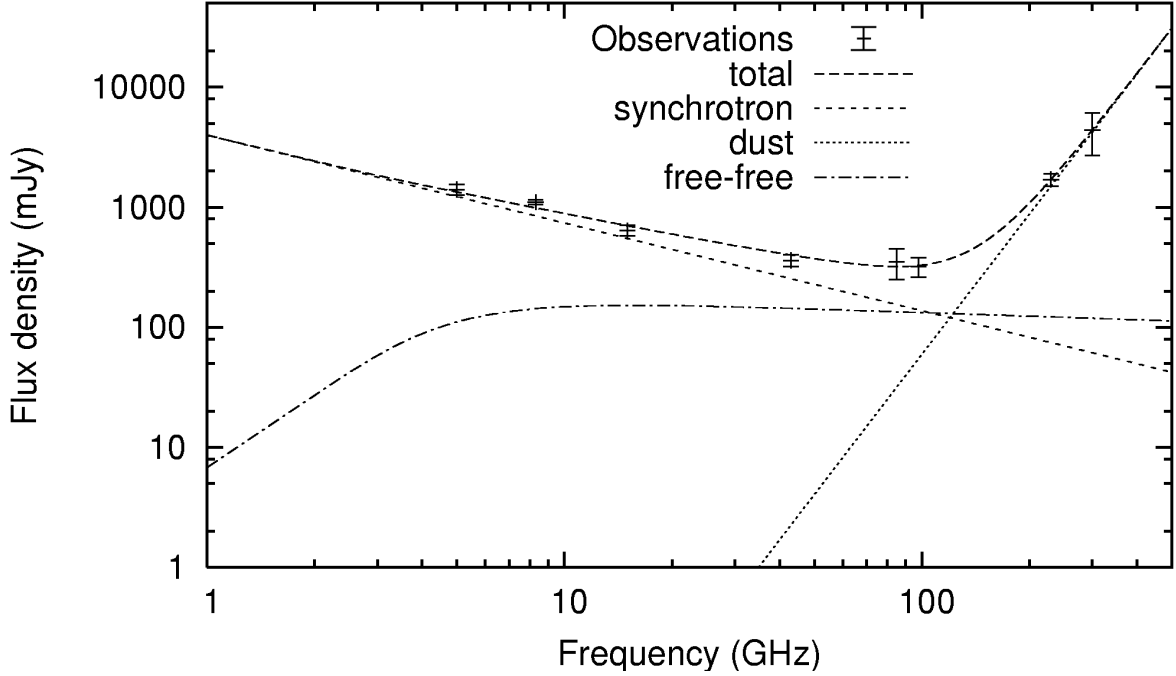}  
\caption{ Integrated radio continuum emission measured toward  
the nuclear $30'' $ region of NGC 253 at 5, 15 GHz (Turner \& Ho 1983), 43 GHz (this work), 85 GHz  
(Carlstrom et al. 1990), 98 GHz (Peng et al. 1996), 230 GHz (Krugel et al.~1990) and 300 GHz (Chini et al. 1984).   
The contributions to the continuum flux density are:  
free-free (S$_{free-free} \propto \nu^{-0.1}$),   
synchrotron (S$_{synchrotron} \propto \nu^{-0.7}$)   
and dust (S$_{dust} \propto \nu^{3.9}$) emission which are shown as indicated.}  
\label{f8}  
\end{figure}

In Table~3, we summarize the 43 GHz continuum flux density measurement along with previous measurements   
($5-300$~GHz) for the central $30''$ region of NGC 253.
These flux density values have been used to determine the relative contributions from free-free,
synchrotron and dust emission.  
At frequencies $<50$~GHz, the relative contributions of free-free and non-thermal  
emission are dominant compared to the thermal dust emission.  
However, at frequencies $>50$~GHz, the dust contribution is more significant.  
The estimated contribution of the thermal free-free emission at 43~GHz   
is $\sim 140$~mJy, while the non-thermal emission accounts for $\sim 220$ mJy.  
These values were obtained assuming that the thermal continuum free-free   
flux density shows S$_{free-free} (\nu) \propto \nu^{-0.1}$. 
Using the observed flux density measurements of the radio continuum in the range of $5-98$~GHz, and
following the procedure used by Turner \& Ho~(1983), the spectral index for the non-thermal 
emission is $\alpha_{synchrotron}=-0.73 \pm 0.06$. After subtracting the free-free and non-thermal emission
from the total continuum emission over the $5-300$~GHz frequency range, we obtain the
spectral index value for the dust emission $\alpha_{dust} = 3.9 \pm 0.2$.
Figure~\ref{f8} shows the contribution from thermal  
free-free, synchrotron and thermal dust emission along with the total  
radio continuum over the frequency range 5 to 300~GHz.
The thermal free-free radio continuum flux density at 43~GHz may be used to  
estimate the ionization rate (N$_{Lyc}$) from NGC 253 using  
\citep{Sch69,Ro80},  
  
\begin{equation}  
\frac{N_{Lyc}}{s^{-1}}=9.0\times 10^{43} \bigg( \frac{S_{C-th}}{mJy}  
\bigg) \bigg( \frac{T_{e}}{10^{4} K} \bigg)^{0.35} \bigg(  
\frac{\nu}{4.9~GHz} \bigg)^{0.1} \bigg( \frac{D}{kpc} \bigg)^2,  
\end{equation}  
  
\noindent where $T_{e}$ is the electron temperature, $\nu$ is the  
frequency and D is the distance to NGC~253. The estimated ionizing  
flux at 43~GHz, assuming T$_e=10^4$~K and D$=2.5$~Mpc is $\sim 7 \times  
10^{52}$~s$^{-1}$. This N$_{Lyc}$ value is a factor of $\sim 5$ lower  
than that estimated by Puxley et al.~(1997), using the H40$\alpha$ line  
flux density and assuming local thermodynamical equilibrium (LTE).
However, the RRL H40$\alpha$ line emission mainly arises from stimulated emission, 
implying a smaller value of the SFR.  
On the other hand, the star formation rate  
($\Psi_{OB}$) can also be estimated using the relation N$_{Lyc}= 5.4  
\times 10^{52} \times \Psi_{OB}$~s$^{-1}$ \citep{An00},
obtained assuming a mass range of $1-100$~M$_{\odot}$ in the  
Miller-Scalo initial mass function (IMF). The total SFR derived in the  
nuclear regions of NGC 253 is thus $\sim 1.3$~M$_{\odot}$~yr$^{-1}$.  
  
The emission in the RRLs H53$\alpha$ and H92$\alpha$ as well as the  
radio continuum at 43 and 8.3 GHz were modeled for the two continuum  
components (NE and the SW) of NGC 253 observed at an angular resolution of
$1\rlap.{''}5 \times 1\rlap.{''}0$ (P.A.$=0^{\circ}$). 
The line and continuum flux densities  
were measured over regions where both the H92$\alpha$ and H53$\alpha$
are detected.
The models consist of a collection of two families of HII  
regions, each with different electron densities. The  
electron density ranges that were explored are n$_e =10^2 -  
10^4$~cm$^{-3}$ and n$_e =10^4 - 10^7$~cm$^{-3}$, for the low- and  
high-density gas components, respectively. The contribution from  
non-thermal synchrotron emission that arise from SNR and the possible  
AGN and free-free emission from HII regions were also considered.  
The RRL emission has been computed considering that the population of  
the atomic levels deviates from local thermodynamic equilibrium  
(LTE). The non-LTE effects result in both internally and externally  
stimulated line emission. The formalism used to compute the RRL and  
radio continuum flux densities at each frequency follows that  
used by \citet{RoR04}. In these models, the electron density (n$_e$),  
temperature (T$_e$), size (s$_o$) and the number of HII regions  
(N$_{HII}$) are free parameters. In order to reduce the number of  
free parameters, the size of each HII region has been assumed to be  
a function of the electron density and the number of ionizing photons  
emitted by the embedded O star by $[$s$_o$~n$_e^{2/3}]= U$, where $U$  
is a function of the spectral type of the star \citep{Pa73}.
In these models, early-type O7 stars have been used as the  
source of the ionizing continuum flux for each HII region. The  
results obtained for n$_e$ and N$_{HII}$ were constrained by the  
observed line and continuum flux densities as well as the volume of  
the total line emission, assuming spherical HII regions.  
  
The models suggest that the thermally ionized gas in the  
NE component consists of a collection of extended ($\sim 1-4$~pc)  
low-density ($\sim 10^2-10^3$~cm$^{-3}$) HII regions and compact  
($\sim 0.01-0.06$~pc) high-density ($\sim 10^5-10^6$~cm$^{-3}$) HII  
regions. In order to reproduce the observations, the mass of ionized 
gas in the low-density component must be a factor of $\sim 10^{3}$
larger than the mass of ionized gas in the high-density component.
The ionized gas in the SW region also consists of low- and  
high-density HII regions, characterized by n$_e \simeq 10^2$~cm$^{-3}$  
and n$_e \simeq 10^5-10^6$~cm$^{-3}$.
Table~4 lists the physical parameters of these HII regions: the electron temperature (T$_e$), 
electron density (n$_e$), size, emission measure (EM), mass of ionized gas,
continuum optical depth ($\tau_c$) at 8.3 and 43 GHz, departure coefficients (b$_n$ and $\beta_n = 1-(k T_e/h\nu_L) d$~ln~$b_n/dn$, where $n$ 
is the quantum number ) 
for the H53$\alpha$ and H92$\alpha$ RRLs, the contribution of free-free emission and the Lyman continuum photons rate.
The second and third columns list these parameters for the low and high-density HII regions on the NE component.
The fourth and fifth columns list the corresponding parameters for the SW component.
Based on these models, the RRL H92$\alpha$ arises mainly as  
externally stimulated line emission from the extended low-density  
($\sim 10^3$~cm$^{-3}$) HII regions, with $\beta_n < -20$. 

The thermal free-free contribution from both the low- and high-density component  
to the total observed continuum emission ranges from $\sim 30\%$ to $\sim 80\%$. 
 The total mass of thermally ionized gas in each component (NE and the SW) is $\sim 10^3$~M$_{\odot}$.  
The Lyman continuum emission rate of $5.9 \times 10^{52}$~s$^{-1}$ obtained from the RRL emission models (see Table~4)
is consistent with $7 \times 10^{52}$~s$^{-1}$, obtained from the continuum emission models (shown in Figure~8).
The Lyman continuum emission rates for the NE and SW components (listed in Table~4) 
were obtained only for regions where H53$\alpha$ line emission was detected. 
On the other hand, the value of $7 \times 10^{52}$~s$^{-1}$ was estimated by 
integrating over the central $30''$ region, which explains the slightly different results.
The results obtained for the low-density gas component  
for the NE component of NGC 253 are in agreement with the results obtained  
by \citet{Mo05} for the 15~pc region, observed with higher angular  
resolution observations ($\sim 0\rlap.{''}3$) and using VLA H166$\alpha$, H92$\alpha$ and  
H75$\alpha$ line data. 
Even though the mass contribution of the high density HII regions
is only $\sim 1\%$ of the total HII mass, this feature contributes $\sim 40\%$ of the 
total H53$\alpha$ line emission.

As noted before by \citet{Mo05}, high angular resolution  
observations are essential in the determination of the physical parameters of both low and high-density HII regions.
In addition to high angular resolution, high frequency observations (e.g. H53$\alpha$) 
are required to determine the properties of the high-density ($\sim~10^5$~cm$^{-3}$)  HII regions. 
Non-LTE effects are important, leading to an enhancement of the line emission by a factor of $\sim 2$.
Thus, the results obtained assuming  LTE conditions overestimate the number of HII regions (and also the SFR) that must exist in NGC~253.
  
\subsection{Kinematics}  
  
The previous H92$\alpha$ line observations at $1\rlap.{''}8\times 1\rlap.{''}0$ angular resolution  
\citep{An96} revealed velocity gradients of $\sim 11$~km~s$^{-1}$~arcsec$^{-1}$ and   
$\sim 18$~km~s$^{-1}$~arcsec$^{-1}$ along the major axis and minor axis, respectively.  
A qualitative comparison of the velocity field observed in the H53$\alpha$ line and the previously reported   
H92$\alpha$ line velocity field reveals kinematical behaviors that are consistent,  
i.e. the regions with red-shifted and blue-shifted ionized gas coincide (see Figs.~\ref{f5} and \ref{f6}).
The coincidence of the red- and blue-shifted regions  
in the H53$\alpha$ and H92$\alpha$ velocity fields implies that   
both the low- and high-density ionized gas components rotate in the same sense.   

\begin{figure}[!ht]  
\epsscale{0.8}  
\plotone{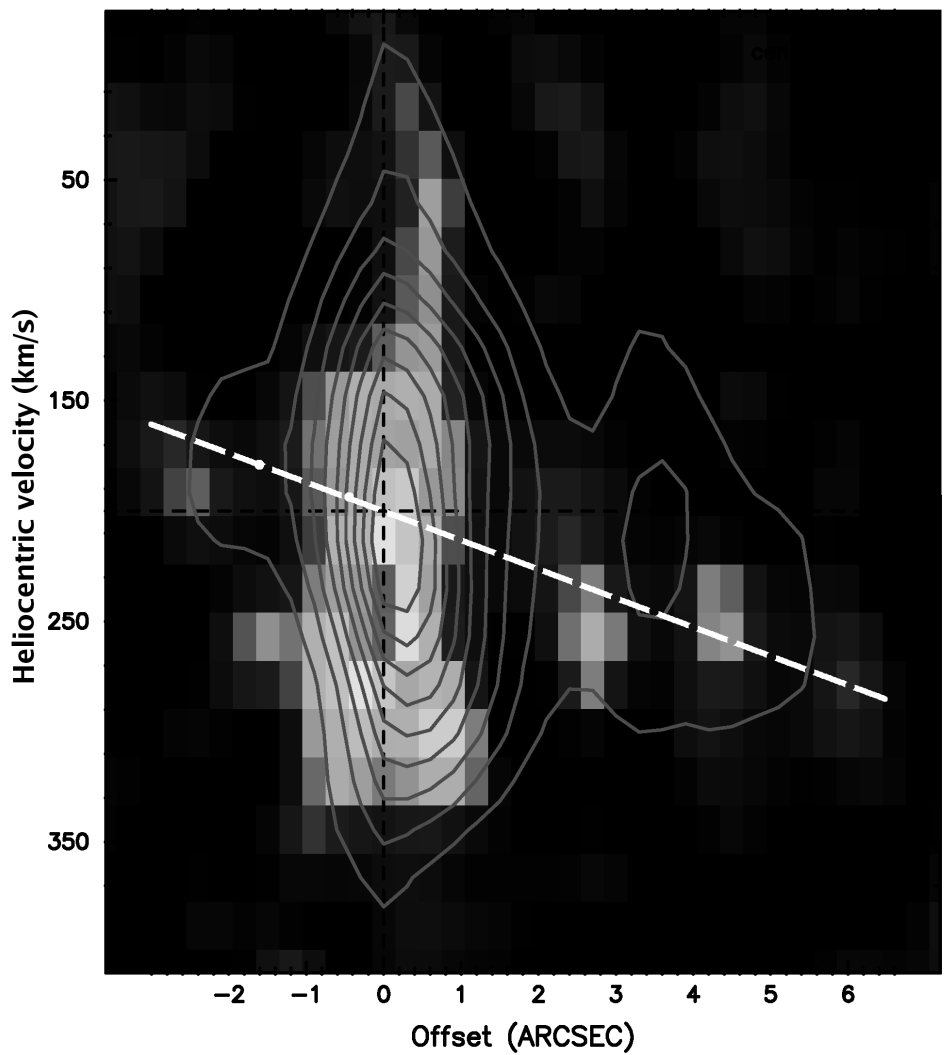}  
\caption{Position-velocity comparisons along the major axis (P.A.$=52^{\circ}$) of NGC 253.  
The H53$\alpha$ RRL PV diagram (color scale) was constructed from the data with  
angular resolution of $1\rlap.{''}5 \times 1\rlap.{''}0$, P.A.$=0^{\circ}$  
and velocity resolution of 44~km~s$^{-1}$. The color scale covers the range $0.64-6.4$~mJy~beam$^{-1}$.  
The H92$\alpha$ RRL PV diagram (contours) was constructed from the  
data of Anantharamaiah \& Goss (1996) with angular resolution of $1\rlap.{''}5 \times 1\rlap.{''}0$, P.A.$=0^{\circ}$  
and velocity resolution 56~km~s$^{-1}$. Contours are 10\%, 20\%, \ldots, 90\%  
of the peak intensity 3.9~mJy/beam. The white dashed line shows the fitted velocity gradient  
of $11$~km~s$^{-1}$~arcsec$^{-1}$.}  
\label{f9}  
\end{figure}

\begin{figure}[!ht]  
\plotone{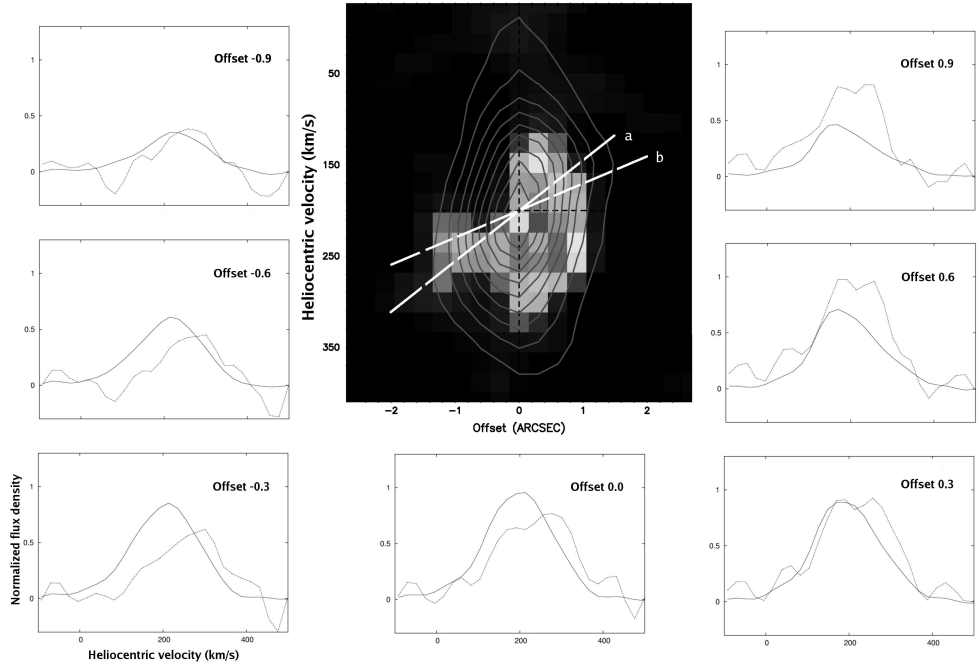}  
\caption{Position-velocity diagrams along a P.A.$=-45^{\circ}$ in NGC 253.  
The H53$\alpha$ RRL PV diagram (color scale) was constructed from the data with  
angular resolution of $1\rlap.{''}5 \times 1\rlap.{''}0$, P.A.$=0^{\circ}$  
and velocity resolution of 44~km~s$^{-1}$. The color scale is from 0.64 to 6.4~mJy/beam.  
The H92$\alpha$ RRL PV diagram (contours) was constructed from the  
data of Anantharamaiah \& Goss (1996), with angular resolution of $1\rlap.{''}5 \times 1\rlap.{''}0$, P.A.$=0^{\circ}$  
and velocity resolution of 56~km~s$^{-1}$. Contours are 10\%, 20\%, \ldots, 90\% of the peak intensity 3.9~mJy/beam.  
The white dashed lines marked with 'a' and 'b' show the fitted velocity gradients of $42$~km~s$^{-1}$~arcsec$^{-1}$ and  
$25$~km~s$^{-1}$~arcsec$^{-1}$, respectively. The panels show the H53$\alpha$ (blue) and H92$\alpha$   
(red) line profiles for the central $1\rlap.{''}8$. Each panel corresponds to a slice taken  
along the velocity axis of this PV diagram e.g. the panel at 0.0 is a slice along the velocity axis at offset position $0\rlap.{''}0$.}  
\label{f10}  
\end{figure}

In order to compare in detail the kinematics of the ionized gas as observed in the H53$\alpha$  
and the H92$\alpha$ lines, we constructed PV diagrams along the major axis (P.A.$=52^{\circ}$) 
using the task SLICE in GIPSY. 
These PV diagrams are shown in Figure~\ref{f9}; the white line is the resulting fit 
to the velocity gradient of $11$~km~s$^{-1}$~arcsec$^{-1}$.
The same procedure was used to obtain PV diagrams along the P.A.$=-45^{\circ}$, shown in Figure~\ref{f10};
the white lines are the resulting fits to the H92$\alpha$ velocity gradient ($24$~km~s$^{-1}$~arcsec$^{-1}$)
and the H53$\alpha$ velocity gradient ($42$~km~s$^{-1}$~arcsec$^{-1}$).
In Figure~10 we also show the H92$\alpha$ and H53$\alpha$ spectra superimposed at 
different offset positions from the central source (5.79-39.0); these spectra were  
normalized based on the peak line flux densities.
By inspection of the different spectra obtained at the negative offset positions 
(between $-0\rlap.{''}6$ and $-0\rlap.{''}9$ in Figure~10), a relative velocity shift ($\sim 30 \pm 5$~km~s$^{-1}$) is observed
between the peak flux density of the H53$\alpha$ and the H92$\alpha$ lines. 
Based on the line emission models, these two RRLs trace different density components (see section~4.1).  
The small velocity shift between these two RRLs on the NE component suggests that
each density component has slightly different kinematics.
  
\subsubsection{Gaseous bar structure, outflow or an accreted object}  
  
Observations at IR wavelengths have revealed the existence of a gaseous bar in NGC~253 \citep{Sco85}.  
In a bar potential the gas follows two types of orbits, x1 and x2.   
The x1 (bar) orbits are those extended along the  
major axis of the bar and the x2 (anti-bar) orbits are those oriented  
perpendicular to the bar major axis. In the case of NGC 253, the x1  
and x2 orbits would be oriented on the plane of the sky at P.A. of  
$\sim 70^{\circ}$ and $\sim 45^{\circ}$, respectively.   
In the H53$\alpha$ and H92$\alpha$ RRL images (at $1\rlap.{''}5 \times 1\rlap.{''}0$ angular resolution)  
the orientation of the largest velocity gradient is nearly perpendicular to the orientation of
the x2 orbits.
Since the ionized gas on the NE component rotates in an opposite sense compared to the CO
\citep{An96,Das01}, a simple bar potential does not account for the differences observed between the  
velocity fields of the RRLs (H92$\alpha$ and H53$\alpha$) and CO.
A secondary bar inside the primary bar may be invoked to explain the kinematics observed in the center of NGC 253.
However, further observations and modeling are required to investigate the  existence of this secondary bar.

Weaver et al.~(2002) proposed the presence of a starburst-driven nuclear outflow collimated by a dusty torus,
based on X-ray observations of NGC~253.
In this model, the thermally ionized gas in the center of NGC~253 should be distributed in both 
a starburst ring and a starburst-driven outflow (Weaver et al.~2002).
Observations of the RRL H92$\alpha$ toward the starburst galaxy M82 (Rodr\'{\i}guez-Rico et al. 2004),
 have proben that RRLs may be used to study the ionized gas associated with galactic outflows.
The largest velocity gradient observed in the RRL H92$\alpha$ is oriented 
nearly along the minor axis of NGC 253 (P.A.$=-45^{\circ}$).
Based on this orientation and assuming the H92$\alpha$ RRL in the NE component traces the ionized gas in the outflow, 
the receding side of this outflow would be on the NW and the approaching side on the SE.
If this is the case all the observed ionized gas would be tracing the outflow, explaining
the different rotation sense between the CO and the ionized gas. However, it seems unlikely 
that all the ionized gas is associated with the outflow.

\citet{Das01} propose that the kinematics of the ionized gas traced by the H92$\alpha$ RRL can be explained   
if there is an accreted object with mass of $\sim 10^6$~M$_{\odot}$.
The CO gas which traces the galactic disk of NGC~253 is moving in an opposite sense compared to the ionized gas 
that may be associated with the compact object.
Based on the H53$\alpha$ velocity gradient ($\sim 42$~km~s$^{-1}$~arcsec$^{-1}$)   
along the minor axis of the NE component ($\sim 30$~pc), the inferred dynamical mass is $\sim 5 \times 10^6$~M$_{\odot}$.  
This mass estimate is consistent with that of the accreted object proposed by \citet{Das01}.  
The existence of a compact object is further supported by the higher angular resolution  
($0\rlap.{''}36 \times 0\rlap.{''}21$) H92$\alpha$ observations (Figure~7),
revealing a larger velocity gradient ($\sim 110$~km~s$^{-1}$~arcsec$^{-1}$, at P.A.$\simeq -45^{\circ}$)   
over the central $\sim 0\rlap.{''}6$ (7~pc).   
This H92$\alpha$ velocity gradient of $\sim 110$~km~s$^{-1}$~arcsec$^{-1}$ implies a dynamical mass of $\sim 7 \times 10^6$~M$_{\odot}$,  
similar to the mass determined from the $1\rlap.{''}5 \times 1\rlap.{''}0$ angular resolution  
observations of the RRLs H53$\alpha$ and H92$\alpha$.
The $\sim 7 \times 10^6$~M$_{\odot}$ dynamical mass is based on observations over a region a factor of three times smaller than  
that observed in the $1\rlap.{''}5 \times 1\rlap.{''}0$ angular resolution images.  

The estimated dynamical mass ($\sim 7 \times 10^6$~M$_{\odot}$) for the nuclear region of NGC 253   
is comparable to that of the compact source at the nucleus of   
our galaxy ($\sim 4 \times 10^{6}$~M$_{\odot}$, Ghez et al.~2005).  
This mass estimate of $\sim 7 \times 10^6$~M$_{\odot}$ for the central region of NGC 253 could exist in the form of a large   
number of stars combined with ionized gas and may also contain an AGN.
If the Lyman continuum photons rate ($\sim 7 \times 10^{52}$~s$^{-1}$) is mainly due to O5 stars, each emitting $\sim 5 \times 10^{48}$~s$^{-1}$,
then there must be $\sim 10^4$ O5 stars in the NE component.
Using Salpeter's initial mass function and a mass range of $0.1-100$~M$_{\odot}$,
the total mass in stars in the NE component is $\sim 10^7$~M$_{\odot}$.
The ionized gas could be gravitationally bounded by the stars and consequently the 
black hole mass would be $\leq 10^7$~M$_{\odot}$.
The existence of an AGN has been proposed from radio continuum  
observations \citep{Tu85,Ul97} and RRL H92$\alpha$ observations \citep{Mo02}.
Radio continuum observations (1.3 to 20 cm), reveal that the  
strongest radio source 5.79-39.0 has a brightness temperature $> 40,000$~K at 22~GHz and is unresolved ($<1$~pc, Ulvestad and Antonucci 1997).
Broad ($> 100$~km~s$^{-1}$) H$_2$O maser line emission is observed toward the  
nuclear regions supporting the existence of a massive object in the center of NGC~253 \citep{Na95}.   

In order to account for the different kinematics observed for the ionized and 
the molecular gas, two possible scenarios can be proposed:
(1) a dense object that is accreted into the nuclear region of NGC 253,
(2) the ionized gas is moving in a starburst-driven outflow and/or
(3) a secondary bar exists within the primary bar, as proposed for other galaxies \citep{Fr93}.
The accreted object model is supported by the high angular resolution H92$\alpha$ observations;
the estimated dynamical mass of $\sim 7 \times 10^6$~M$_{\odot}$ is concentrated in a $\le 7$~pc region and the
ionized gas traced by the RRLs moves in the opposite direction compared to the larger scale CO.
In the RRLs H53$\alpha$ and H92$\alpha$, we find no evidence that confirms the existence of a secondary bar.
An S-shape in the velocity field is characteristic of a bar \citep{An96}. Thus, for a secondary bar a second 
S-shaped pattern would be observed in the velocity field which is not appreciated in the H92$\alpha$ velocity structure (see Figure~7).
However, this scenario cannot be ruled out and higher angular and spectral resolution observations are
necessary to discern between these three models.

\section{CONCLUSIONS.}   
  
The H53$\alpha$ RRL and radio continuum at 43 GHz were observed at high angular  
resolution ($1\rlap.{''}5 \times 1\rlap.{''}0$) towards NGC 253.  
We have also reanalyzed previous observations of the RRL H92$\alpha$ made at an angular  
resolutions of $1\rlap.{''}5 \times 1\rlap.{''}0$ \citep{An96} and   
$0\rlap.{''}36 \times 0\rlap.{''}21$ \citep{Mo02}.  

Based on the 43 GHz radio continuum flux density 
and previous measurements at lower and higher frequencies, we have estimated the contribution
from free-free emission ($\sim 100$~mJy at 43~GHz). Using this value for the free-free emission,
the derived SFR in the nuclear region of NGC~253 $\sim 1.3$~M$_{\odot}$~yr$^{-1}$.
The RRLs (H53$\alpha$ and H92$\alpha$) and radio continuum (at 43 and 8.3 GHz) emission have been modeled  
using a collection of HII regions.
Based on the models, the RRL H53$\alpha$ enables us to trace the compact 
($\sim 0.1$~pc) high-density ($\sim 10^5 - 10^6 $~cm$^{-3}$) HII  
regions in NGC 253. The total mass of high-density ionized gas in the central 18~pc is $\sim 10^3$~M$_{\odot}$.   
A large velocity gradient ($\sim 42$~km~s$^{-1}$~arcsec$^{-1}$, P.A.$\sim -45^{\circ}$) is observed in the H53$\alpha$ line.  
The orientation and amplitude of the velocity gradients (angular resolution $\sim 1\rlap.{''}2$) 
derived using the H53$\alpha$ and H92$\alpha$ lines agree. 
The high angular resolution observations ($0\rlap.{''}36 \times 0\rlap.{''}21$)  
of the H92$\alpha$ line reveal a larger velocity gradient ($\sim 110$~km~s$^{-1}$~arcsec$^{-1}$); this large velocity gradient
implies a dynamical mass on the NE component ($\leq 7$~pc) of $\sim 7 \times 10^6$~M$_{\odot}$, supporting the
existence of an accreted compact object.  

The orientation of the H53$\alpha$ and H92$\alpha$ velocity gradients does not agree with the CO kinematics.
 The different kinematics observed over a larger region of the
disk of NGC 253 also suggests the existence of an accreted object.
The derived dynamical mass ($\sim 7 \times 10^{6}$~M$_{\odot}$) can be accounted for by a large stellar density and/or 
the presence of an AGN.  
The star formation activity in NGC~253 may be the result of a merger  
process of NGC~253 and this proposed massive compact object.   
  
The National Radio Astronomy Observatory is a facility of the National  
Science Foundation operated under cooperative agreement by Associated  
Universities, Inc. CR and YG acknowledge support from UNAM and  
CONACyT, M\'exico.

\clearpage  
  
\begin{deluxetable}{cc}  
\tablecolumns{2}    
\tablewidth{0pc}   
\tablecaption{Observing parameters for NGC~253 using the VLA.}  
\tablehead{\colhead{Parameter} & \colhead{H53$\alpha$ RRL (43 GHz)}}
\startdata
Right ascension (J2000) \dotfill & 00$^h$47$^m$$33\rlap.{^s}$18\\
Declination (J2000)\dotfill & $-25 ^{\circ} 17' 17\rlap.{''}0$\\ 
Angular resolution continuum\dotfill & $1\rlap.{''}5 \times 1\rlap.{''}0$, P.A.$=0^{\circ}$ \\   
Angular resolution line\dotfill & $1\rlap.{''}5 \times 1\rlap.{''}0$, P.A.$=0^{\circ}$ \\   
On-source observing duration (hr)\dotfill & 8  \\   
Bandwidth (MHz)\dotfill & 125\\ Number of spectral channels\dotfill & 38\\   
Center V$_{Hel}$ (km~s$^{-1}$)\dotfill & 200   \\  
Velocity coverage (km~s$^{-1}$)\dotfill & 1000 \\   
Velocity resolution (km~s$^{-1}$)\dotfill & 44 \\   
Amplitude calibrator\dotfill & J$0137+331$\\   
Phase calibrator\dotfill & J$0120-270$\\   
Bandpass calibrator\dotfill & J$0319+415$\\   
RMS line noise per channel (mJy/beam)\dotfill & 0.65\\   
RMS, continuum (mJy/beam)\dotfill & 0.2\\  
\enddata  
\end{deluxetable}  
  
\clearpage    

\begin{deluxetable}{lccccccccccc}  
\tabletypesize{\scriptsize}  
\rotate
\tablecolumns{12}
\tablewidth{0pc}
\tablecaption{H53$\alpha$ and H92$\alpha$ line parameters for NGC~253. }  
\tablehead{ & & & \colhead{H53$\alpha$} & & & & & &\colhead{H92$\alpha$} \\   
\cline{2-6} \cline{8-12} \\   
\colhead{} & \colhead{S$_{C43}$\tablenotemark{a}} & \colhead{S$_L$} & \colhead{$\Delta V_{FWHM}$} & \colhead{V$_{Hel}$} &  
\colhead{$S_L \cdot \Delta V_{FWHM}$} & & \colhead{S$_{C8.3}$\tablenotemark{a}} & \colhead{S$_L$} & \colhead{$\Delta V$} & \colhead{V$_{Hel}$} &  
\colhead{$S_L \cdot \Delta V_{FWHM}$} \\   
\colhead{Feature} & \colhead{(mJy)} & \colhead{(mJy)} & \colhead{(km~s$^{-1}$)} &  
\colhead{(km~s$^{-1}$)} & ($\times 10^{-20}$~W~m$^{-2}$) & & \colhead{(mJy)} & \colhead{(mJy)} &  
\colhead{(km~s$^{-1}$)} & \colhead{(km~s$^{-1}$)} & ($\times 10^{-22}$~W~m$^{-2}$) }   
\startdata
NGC 253 NE & $134 \pm 5$ & $9.0 \pm 0.5$ & $200 \pm 10$ & $220 \pm 5$ & $0.25 \pm 0.01$ & &  
$278 \pm 10$ & $4.5 \pm 0.1$ & $190 \pm 5$ & $200 \pm 2$ & $2.4 \pm 0.2$ \\   
NGC 253 SW & $24 \pm 2$ & $2.8 \pm 0.3$ & $130 \pm 20$ & $215 \pm 10$ & $0.04 \pm 0.01$ & & $40  
\pm 4$ & $0.7 \pm 0.1$ & $160 \pm 15$ & $220 \pm 10$ & $0.3 \pm 0.03$ \\   
NGC 253 Total & $360 \pm 20$ & $21 \pm 2$ & $230 \pm 20$ & $210 \pm 10$ & $0.69 \pm  
0.09$ & & $ 605 \pm 10 $ & $ 9.0 \pm 0.5$ & $ 190\pm 10 $ & $ 206 \pm 4 $ & $ 4.7 \pm 0.5 $ \\   
\enddata
\tablenotetext{a}{Continuum emission was obtained by integrating over the region where both RRLs H53$\alpha$ and H92$\alpha$ were detected.}
\end{deluxetable}  

\clearpage  

\begin{deluxetable}{ll}  
\tablecolumns{2} \tablewidth{0pc}   
\tablecaption{Observed continuum flux densities for NGC~253\tablenotemark{a}. }  
\tablehead{ \colhead{Frequency} & \colhead{Flux density}\\  
\colhead{(GHz)} & \colhead{(mJy)}}   
\startdata   
5   & $ 1400 \pm 70$ \tablenotemark{b}\\   
8.3 & $ 1100 \pm 25$ \tablenotemark{c}\\   
15  & $ 640  \pm 30$ \tablenotemark{b}\\   
43  & $ 360  \pm 20$ \tablenotemark{d}\\   
85  & $ 350  \pm 50$ \tablenotemark{e}\\   
99  & $ 320  \pm 30$ \tablenotemark{f}\\   
230  & $ 1700  \pm 200$ \tablenotemark{g}\\   
300  & $ 4400  \pm 1700$ \tablenotemark{h}\\   
\enddata
\tablenotetext{a}{Integrated flux density measured over the inner $\sim~30''$ region.}
\tablenotetext{b}{Turner \& Ho~(1983).}
\tablenotetext{c}{Mohan et al.~(2002).}
\tablenotetext{d}{This  work.}
\tablenotetext{e}{Carlstrom et al.~(1990).}
\tablenotetext{f}{Peng et al.~(1996).}
\tablenotetext{g}{Krugel et al.~(1990).}
\tablenotetext{h}{Chini et al.~(1984).}
\end{deluxetable}  

\clearpage  
  
\begin{deluxetable}{lcccccc}  
\tablecolumns{7} \tablewidth{0pc} \tablecaption{Results from models using a collection of  
HII regions for  NGC~253. }  
\tablehead{   	    & \multicolumn{2}{c}{NE region} & & \multicolumn{2}{c}{SW region} \\
\colhead{Parameter} & \colhead{Low density\tablenotemark{a}} & \colhead{High density\tablenotemark{a}} &
		    & \colhead{Low density\tablenotemark{a}} & \colhead{High density\tablenotemark{a}} }   
\startdata   
T$_e$ (K)	        & $7500 \pm 2500$ 		& $7500 \pm 2500$ 		& & $ 5000 \pm 1000$ 	        & $ 5000 \pm 1000$ \\   
n$_e$ (cm$^{-3}$)	& $1.6 \pm 1.4 \times 10^3$ 	& $5.5 \pm 4.5 \times 10^5$ 	& & $ 200 \pm 100$ 	        & $ 6.5 \pm 3.5 \times 10^5 $\\
Size (pc)		& $2.4 \pm 1.7$ 		& $0.04  \pm 0.03$ 		& & $ 4.5 \pm 1.5$ 	        & $ 0.02 \pm 0.01$ \\ 
EM (cm$^{-6}$~pc)	& $3\pm2 \times 10^6$ 		& $50 \pm 45 \times  10^8$ 	& & $ 1.5\pm 1 \times 10^5 $      & $ 6 \pm 4 \times 10^9 $ \\ 
M$_{HII}$  (M$_{\odot}$)& $3 \pm 2 \times 10^3 $ 	& $7 \pm 6 $			& & $ 5 \pm 3  \times 10^3 $      & $ 0.4 \pm 0.3 $ \\ 
$\tau_C$ (8.3~GHz) 	& $6 \pm 5 \times 10^{-3}$ 	& $32 \pm 30$			& & $ 6 \pm 5 \times  10^{-4}$    & $ 60 \pm 40 $\\ 
$\tau_C$ (43~GHz) 	& $6 \pm 5 \times 10^{-5} $ 	& $1.0 \pm 0.9$ 		& & $ 4 \pm 3 \times  10^{-5} $   & $ 1.9 \pm 1.3 $ \\ 
$b_n$ (H92$\alpha$) 	& $0.881 - 0.973$ 		& $0.997 - 0.999$ 		& & $ 0.819 - 0.887 $ 		& $ 0.998 - 0.999 $ \\
$b_n$ (H53$\alpha$) 	& $0.70 - 0.79$ 		& $0.96 - 0.98$ 		& & $ 0.64 - 0.65 $ 		& $ 0.97 - 0.98 $ \\				 
$\beta_n$ (H92$\alpha$) & $-55 \pm 32$ 			& $-1.25 \pm 0.75$ 		& & $-70 \pm 10$ 			& $-0.15 \pm 0.14$ \\ 
$\beta_n$ (H53$\alpha$) & $-27 \pm 13$ 			& $-10 \pm 3$ 			& & $-13 \pm 2$ 			& $-5 \pm 1$ \\
\cline{2-3}
\cline{5-6}
S$_{ff-43}$\tablenotemark{b} (mJy)	& \multicolumn{2}{c}{$70  \pm 30$}	& & \multicolumn{2}{c}{$19 \pm 2$} \\ 
N$_{Lyc}$ ($\times 10^{52}$~s$^{-1}$) & \multicolumn{2}{c}{$5 \pm 2$} 		& & \multicolumn{2}{c}{$0.9 \pm 0.1$} \\  
\enddata 
\tablenotetext{a}{Both components were used to fit the continuum emission at  
8.3 and 43 GHz as well as the flux densities in the RRLs H92$\alpha$  and H53$\alpha$.}  
\tablenotetext{b}{Free-free continuum flux density at 43~GHz obtained summing the contributions from the low and high-density components.}
\end{deluxetable}  
  
\clearpage  
  
\clearpage  
  
\clearpage  

\clearpage  

\clearpage  

\clearpage  

\clearpage  

\clearpage  
  
\clearpage    
  
\clearpage  

\end{document}